\documentclass[twocolumn,showpacs,preprintnumbers,amsmath,amssymb,prb,superscriptaddress]{revtex4}

\usepackage{epsfig}

\def\sigbar{{\bar{\sigma}}}
\newcommand{\stat}{\text{stat}}
\newcommand{\sgn}{\text{sgn}}

\renewcommand{\Im}{\operatorname{Im}}

\begin{document}

\title{Transient dynamics of the Anderson impurity model out of equilibrium}

\author{T.~L.~Schmidt}
\affiliation{Departement Physik, Universit\"at Basel, Klingelbergstr.~82, 4056 Basel, Switzerland}

\author{P.~Werner}
\affiliation{Columbia University, 538 West 120th Street, New York, NY 10027, USA}
\affiliation{Theoretische Physik, ETH Zurich, 8093 Zurich, Switzerland}

\author{L.~M\"uhlbacher}
\affiliation{Physikalisches Institut, Universit\"at Freiburg, Hermann-Herder-Str.~3, 79104 Freiburg, Germany}

\author{A.~Komnik}
\affiliation{Institut f\"ur Theoretische Physik, Universit\"at Heidelberg, Philosophenweg 19, 69120 Heidelberg, Germany}

\date{\today}

\begin{abstract}
We discuss the transient effects in the Anderson impurity model
that occur when two fermionic continua with finite bandwidths are
instantaneously coupled to a central level. We present results for
the analytically solvable noninteracting resonant level system
first and then consistently extend them to the interacting case
using the conventional perturbation theory and recently developed
nonequilibrium Monte Carlo simulation schemes. The main goal is to
gain an understanding of the full time-dependent nonlinear
current-voltage characteristics and the population probability of
the central level. We find that, contrary to the steady state, the
transient dynamics of the system depends sensitively on the
bandwidth of the electrode material.
\end{abstract}


\maketitle

\section{Introduction}
The Anderson impurity model (AIM) has been introduced in the early
1960s to describe conduction electrons interacting with a magnetic
atom and since then continues to attract the attention of
condensed matter physicists.\cite{PhysRev.124.41}
Despite some notable exceptions,\cite{wyatt} for quite a long time mainly the zero-bias anomaly as well as other equilibrium properties, which can be extracted from the exact analytical solution via the Bethe ansatz approach, have been in the focus of the theoretical research.\cite{appelbaum,andersonX,tsvelick,andrei} It was only with the advent of nanotechnology that the investigation of nonequilibrium properties received a boost, as it became possible not only to directly manufacture structures which are adequately described by the AIM, but also to investigate their nonequilibrium properties under well controlled
parameters.\cite{goldhaber-gordon,kouwenhoven,weisvonklitzing}

However,
even in the time-independent steady-state case the analysis of the
nonequilibrium situation turns out to be rather difficult.
Despite a large number of works employing various perturbative and
renormalization group techniques
(e.g.~Refs.~[\onlinecite{PhysRevLett.70.2601,PhysRevB.55.3003,konig,
rosch, PhysRevB.58.5649,oguriX, PhysRevB.69.153102, Thorwart}]), or even
attempts at solving
the problem analytically,\cite{KSL} there is no solution which
unifies all known details.

Even more difficult is the case of the ``preparative'' nonequilibrium,
i.e.~the time evolution of the system from some initial preparation towards
its steady state under a finite external voltage bias. For the first time
this problem has been discussed in Ref.~[\onlinecite{jauho1}], where a
solution for the wide flat band (WFB) limit was derived. However, the
assumption of an infinitely wide band leads to the rather unphysical
prediction of a displacement current which instantaneously jumps
to a finite value immediately after switching on the tunneling.

The transient nonequilibrium dynamics of a strongly
interacting quantum dot which is suddenly brought into the Kondo
regime, has been investigated using approximative
schemes.\cite{plihal05,goker07} Moreover, the band structure
effects on the time evolution of noninteracting nanoscale devices
have been investigated in [\onlinecite{maciejko06}].

 However, the combined effect of interaction and finite
 bandwidth, both of which can be described within the framework
 of the AIM, have not yet been considered. In this article we attempt
 to address this issue by  means of perturbation theory in the case
 of weak interactions and Monte Carlo (MC) simulations for moderate
 to strong interactions. Such results are not only interesting for future
 experiments. In view of recent attempts to use the integrability methods
 to understand the nonequilibrium properties of quantum impurity models
 it is important to develop and test numerical schemes which are able
 to generate reliable results for any parameter constellation.
 \cite{PhysRevLett.90.246403,mehta:216802,boulat:033409}

The structure of this paper is as follows. After introducing the
system under consideration in Section \ref{model}, we start with a
resonant level model which maps onto the noninteracting AIM.
Because the corresponding Hamilton operator is quadratic in the
fermionic fields, the dynamics of the system can be investigated
by analytic means at any parameter constellation even for a model
with arbitrary band structure of the electrodes (cutoff schemes).
The basis of our solution is the integral equation for the
impurity retarded Green's function (GF), which we derive next. It
is then used in Section \ref{basicthings} for the calculation of
the time-dependent impurity population function $n(t)$ as well as
for the expectation value of the transient current. Here we not
only consider the simplest case of a WFB structure of the leads,
but also more realistic models taking into account bandwidth
effects. Section \ref{perttheory} is devoted to the analysis of
the transient dynamics of an interacting system. Using
perturbation theory in interaction strength $U$ we identify the
leading order effects in that limit. A treatment of arbitrary
interaction strengths is best accomplished with the help of a
dedicated nonequilibrium Monte Carlo (MC) scheme, which is
presented in Section \ref{MC}.

\section{Model and observables}                    \label{model}
The AIM Hamiltonian usually consists of four
contributions,\cite{PhysRev.124.41}
\begin{eqnarray} \label{Hamiltonian}
H = H_{\text{dot}} + H_0[\psi_{R,L}] + H_T + H_U \, .
\end{eqnarray}
$H_{\text{dot}}$ describes two spin-degenerate fermionic levels
with energy $\Delta$ (which we later shall also call ``dot''),
\begin{eqnarray}
 H_{\text{dot}} = \Delta \sum_{\sigma=\uparrow, \downarrow} \,
d^\dag_\sigma \, d_\sigma \, .
\end{eqnarray}
It is coupled to two fermionic continua -- electrodes on the left
and right sides. Each of these is modeled by a field operator
$\psi_\alpha(x)$ (where $\alpha = L,R$) and the corresponding
Hamiltonians $H_0[\psi_\alpha]$, whose precise shape we shall
discuss in a moment. The operator $H_T$ is responsible for the
particle exchange between the dot and the electrodes and is given
by a simple local tunneling term of the form
\begin{eqnarray}                 \label{HT}
H_T = \sum_{\alpha=R,L} \sum_{\sigma} \gamma_\alpha \left[
\psi_{\alpha\sigma}^\dag(x=0) \, d_\sigma + \text{
h.c.} \right] \, .
\end{eqnarray}
Finally, $H_U$ accounts for the interaction in the system and is
formally implemented as an additional energy cost for the double occupancy
of the dot level,
\begin{eqnarray}                         \label{H_U}
 H_U = U \, d^\dag_\uparrow d_\uparrow \, d^\dag_\downarrow
d_\downarrow \, .
\end{eqnarray}
In general, it is quite difficult to analyze the properties of the
interacting Anderson model at $U \neq 0$, but not impossible. In
fact, at least in equilibrium an exact analytic solution can
be derived via the Bethe ansatz.\cite{tsvelick,andrei} In the
genuine nonequilibrium, when a finite bias voltage is applied
across the dot, the picture is far from complete since as yet no
exact solution exists.

On the other hand, since the tunneling part $H_T$, being only
quadratic in the fermionic operators, is diagonalizable by
elementary methods and the Green's functions (GFs) are accessible
in all parameter regimes, the noninteracting system [aka resonant
level (RL) model] is exactly solvable by elementary
means.\cite{caroli} In this case, the Hamiltonian as well as
expectation values separate into spin-up and spin-down
contributions, so that throughout this and the next Sections we
shall work with spinless operators, recovering the necessary
prefactors after the calculations.

In the case of the initially uncoupled dot GFs we have to deal
with two different situations: (i) the dot level is empty,
$n_0=0$, and (ii) the dot is populated by one electron,  $n_0=1$.
Due to the simple structure of $H_{\text{dot}}$, the time
evolution is trivial, $d(t) = d(0) \, \exp( - i \Delta t )$,
immediately leading to the following matrix (Keldysh) GF
[\onlinecite{LLX}],
\begin{widetext}
\begin{eqnarray}
 {\bf D}_0(t) = \left[ \begin{array}{cc}
D_0(t) & D_0^{<}(t) \\
D_0^{>}(t) & \widetilde{D}_0(t)
 \end{array} \right] =
 e^{- i \Delta t} \left[ \begin{array}{cc}
       -i [\Theta(t) (1-n_0) - \Theta(-t) n_0] &  i n_0 \\
       - i (1-n_0) & - i [\Theta(-t)(1-n_0) - \Theta(t) n_0] \\
                        \end{array} \right] \, ,
\end{eqnarray}
\end{widetext}
where $D_0(t)$ and $\widetilde{D}_0(t)$ denote the time-ordered
and anti-time-ordered GFs, respectively. For the retarded and
advanced components we obtain,
\begin{eqnarray}
 D_0^R(t) &=& D_0(t) - D_0^<(t) = - i \Theta(t) e^{- i \Delta t} \, ,
\\ \nonumber
 D_0^A(t) &=&  D_0^<(t) - \widetilde{D}_0(t) =  i \Theta(-t) e^{- i
\Delta t} \, .
\end{eqnarray}
The Hamiltonian for the electrode electrons can generally be written as
(for $\alpha = R,L$)
\begin{eqnarray}    \label{H0electrode}
 H_0[\psi_\alpha]= \sum_{\bf k} \epsilon_{\bf k} \, \psi_{\alpha {\bf k}}^\dag
 \psi_{\alpha {\bf k}} \, ,
\end{eqnarray}
implying a trivial time evolution of the field operators. Due to
the local tunneling assumption made in Eq.~(\ref{HT}), coupling to
the leads only involves the operator
\begin{eqnarray}
\psi_{\alpha} (x=0) = \sum_{\bf k} \, \psi_{\alpha {\bf k}} \, .
\end{eqnarray}
Therefore, we only need \emph{local} GFs of the band degrees of
freedom in all subsequent calculations and can suppress the
coordinate variable. For the retarded GF we thus have
\begin{eqnarray}  \label{theRGF}
 g_\alpha^R(t)
 = - i \Theta(t) \int d \omega \rho(\omega) \, e^{- i
 \omega t} \,  ,
\end{eqnarray}
where we have introduced the energy-dependent density of states (DoS)
$\rho(\omega)$. In a similar way, one obtains the full Keldysh matrix
\begin{eqnarray}
 {\bf g}_\alpha(\omega) = i 2 \pi \rho(\omega) \left[
 \begin{array}{cc}
 f_\alpha - 1/2 & f_\alpha \\
-(1-f_\alpha) & f_\alpha - 1/2
\end{array} \right] \, .
\end{eqnarray}
Here, $f_\alpha$ denotes the Fermi distribution function in the
respective electrode $\alpha = L,R$.\footnote{We assume the
electrodes to be large enough in order to have a meaningful
concept of temperature.} The retarded and advanced components are
easily retrieved, $g_\alpha^R(\omega) = - i \pi \rho(\omega)$ and
$g_\alpha^A(\omega) = [g_\alpha^R(\omega)]^* = i \pi
\rho(\omega)$. We would like to point out that the actual
dimensionality of the electrode disappears from the problem during
the transition from Eq.~(\ref{H0electrode}) to (\ref{theRGF}),
since it is completely encoded in the DoS.

The GFs of the coupled system can be found
analytically for arbitrary time dependence $\gamma_\alpha(t) \neq 0$ of the tunneling constant. From now on, we shall concentrate on the case of
sudden switching $\gamma_\alpha(t)= \gamma_\alpha \, \Theta(t)$,
where $\Theta(t)$ is the Heaviside step function. A generalization
to arbitrary time dependence is relatively straightforward. The
way to obtain the necessary Dyson equation is precisely the same
as in the stationary case. The result can be summarized in the
matrix equation\cite{caroli,jauho1} (for $t,t' \geq 0$),
\begin{eqnarray}                 \label{fundDyson}
 {\bf D}(t,t')
& = &
 {\bf D}_0(t-t') +
 \int_{0}^\infty \, d t_1 \,
 \int_{0}^\infty \, d t_2 \nonumber \\
& \times &
 {\bf D}_0(t-t_1) \,
 \boldsymbol{\Sigma}(t_1 - t_2) \, {\bf D}(t_2,t') \, ,
\end{eqnarray}
where the generalized self-energy is defined by
\begin{eqnarray}                    \label{selfenergy}
 \boldsymbol{\Sigma}(t) =  \gamma_L^2 \, {\bf g}_{L}(t) +
 \gamma_R^2 \, {\bf g}_R(t) \, .
\end{eqnarray}
The seemingly complicated structure of Eq.~(\ref{fundDyson}) simplifies considerably for the retarded GF,
\begin{eqnarray}                 \label{retardedDyson}
 D^{R}(t,t')
 &=&
 D^R_0(t-t')
\\ \nonumber
 &+& \int_0^\infty \,
 d t_2 \, K(t,t_2) \, D^R(t_2,t')
 \, ,
\end{eqnarray}
where
\begin{eqnarray}                        \label{thekernel}
 K(t,t_2) &=& \int_0^\infty d t_1 \, D^R_0(t-t_1) \, \Sigma^R(t_1 -
 t_2)
\end{eqnarray}
is the kernel of the integral equation.

The simplest physical quantity to calculate is the time-dependent
dot population $n(t) = \langle d^\dag(t) \, d(t) \rangle$. It is
convenient to rewrite it in terms of the off-diagonal Keldysh GF,
\begin{eqnarray}                          \label{nddef}
 n(t) = - i \, D^{<}(t,t) \, .
\end{eqnarray}
The necessary relation between this function and the already known retarded
GF is provided by\cite{langreth,mahan}
\begin{equation}                 \label{longerformula}
 D^{<} = (1+ G^R \Sigma^R ) \, D_0^{<} \, (1+\Sigma^A \, G^A)
+ G^R
 \, \Sigma^{<} \, G^A \, ,
\end{equation}
where products denote integration over time. This relation is
especially useful for the case of an initially empty dot since
then $D^{<}_0=0$ and only the last term contributes. (A similar
relation can be derived for the counterpart $D^{>}_0$, which would
be useful for the initially populated dot.)

Another important observable is the current through the system.
The canonical way to calculate it is to start from the total
particle number operator,\cite{mahan} e.g.~in the left electrode
$Q_L$, and to use the Heisenberg equation to obtain its time
derivative, which is proportional to the current,
\begin{eqnarray}
 \hat{I}_L &=& - \frac{d Q_L}{d t} =  i [ Q_L, H ]
 \nonumber \\
 &=& i \gamma_L \left(
 \psi_L^\dag \, d - d^\dag \, \psi_L \right) \, .
\end{eqnarray}
The evaluation of its expectation value then leads to mixed
correlation functions of dot and lead operators,
\begin{eqnarray}
 I_L(t) = i \gamma_L \langle \psi_L^\dag(t) \, d(t) \rangle
 - i \gamma_L \langle
 d^\dag(t) \, \psi_L(t) \rangle \, .
\end{eqnarray}
The expectation values entering this formula can be rewritten in
terms of GFs. After placing the time $t$ on the forward Keldysh
branch and performing the contour disentanglement we obtain (we
use the definition $D^K = D^{<} + D^{>}$)
\begin{eqnarray}                  \label{leftcurrent}
 I_L(t) &=& - \frac{\gamma_L^2}{2} \int_0^\infty d t_1 \left\{
 [g_L^A(t,t_1) + g_L^K(t,t_1)] \, D^A(t_1,t)
 \right.  \nonumber \\
 &+& \left.
 g_L^R(t,t_1) \, [D^K(t_1,t) + D^R(t_1,t)]
\right. \\ \nonumber
 &-& \left. [D^A(t,t_1) + D^K(t,t_1)] \, g_L^A(t_1,t)
\right. \\ \nonumber
 &-& \left. D^R(t,t_1) \,
 [ g_L^K(t_1,t) + g_L^R(t_1,t) ]
 \right\}   \, .
\end{eqnarray}
Some products of advanced and retarded GFs vanish as their factors
have time arguments of opposite signs. This simplifies the result considerably,
\begin{eqnarray}
 I_L(t) = I_L'(t) + I_L''(t) \, ,
\end{eqnarray}
where
\begin{eqnarray}                        \label{Ip1}
 I'_L(t) & = &  \frac{\gamma_L^2}{2} \int_0^\infty d t_1 \left[
D^K(t,t_1) \, g_L^A(t_1,t) \right. \nonumber \\
& & \left.
 - g_L^R(t,t_1) \, D^K(t_1,t) \right]
\,  \nonumber \\
& = & - \gamma_L^2 \, \mbox{Re} \int_0^\infty d t_1 \, g_L^R(t,t_1) \,
D^K(t_1,t) \, ,
\end{eqnarray}
and
\begin{eqnarray}                        \label{Ip2}
 I''_L(t)
 = \gamma_L^2 \, \mbox{Re} \int_0^\infty d t_1 \, D^R(t,t_1) \,
g_L^K(t_1,t)  \, ,
 \end{eqnarray}
 after using the antihermiticity of the $g^K$ and $D^K$ GFs.
For the evaluation of these expressions it is convenient to use the
relation $D^K = 2 D^< + D^R - D^A$.

\section{The noninteracting case}             \label{basicthings}

\subsection{Wide flat band limit}
In the stationary situation, the time-translational symmetry of
all quantities entering Eqs.~(\ref{retardedDyson}) and
(\ref{thekernel}) is restored and the integral equation is solved
by a mere Fourier transformation.\cite{caroli} In the dynamic
case, the situation is more complex. We begin with the already
known results obtained in the approximation $\rho(\omega) =
\rho_0$, when the conduction band in the electrodes is assumed to
be of zero curvature over an infinite range of energies. In this
case (we concentrate henceforth on the symmetric coupling case
$\gamma=\gamma_R = \gamma_L$)
\begin{eqnarray}\label{gK_WB}
 g_\alpha^A(t) &=& i \pi \rho_0 \, \delta(t) \, ,
 \nonumber \\
 g_\alpha^R(t) &=& -i \pi \rho_0 \, \delta(t) \, ,
 \\
 K(t,t_2) &=& - \Gamma \, e^{- i \Delta (t-t_2)} \, \Theta(t-t_2) \,
 \Theta(t_2) \, ,
\end{eqnarray}
where $\Gamma = 2 \pi \, \rho_0 \, \gamma^2$. As has been realized
in Refs.~[\onlinecite{PhysRevB.43.2541},\onlinecite{jauho1}], the
integral equation for the retarded GF can then be solved by
iterations. The result has the very appealing form
\begin{eqnarray}\label{DR_flat}
 D^R(t-t') = - i \Theta(t-t') \, e^{ - i \Delta (t-t')} \, e^{ -
 \Gamma (t-t')} \, .
\end{eqnarray}
Gathering all terms we obtain
\begin{eqnarray}           \label{1stresult}
  n(t)
& = &
 \frac{\Gamma}{2 \pi} \int d \omega \, \left[ f_R(\omega) + f_L(\omega)
 \right] \, .
 \nonumber \\
& \times &
 \frac{1 + e^{-2 \Gamma t} - 2 \, e^{-\Gamma t}
 \, \cos\left[ (\omega - \Delta) t \right]}{\Gamma^2 + (\omega -
 \Delta)^2} \, ,
 \nonumber \\
& = &
 n_\stat (1 + e^{-2 \Gamma t}) \\
& - &
 \frac{\Gamma e^{-\Gamma t}}{\pi} \int d \omega \,
 \left[ f_R(\omega) + f_L(\omega) \right]
 \frac{
 \, \cos\left[ (\omega - \Delta) t \right]}{\Gamma^2 + (\omega -
 \Delta)^2} \, . \nonumber
\end{eqnarray}
The asymptotic, steady-state value for the population is given by
\begin{eqnarray}
 n_\stat
& = &
 \frac{\Gamma}{2 \pi}  \int d \omega \, \frac{ f_R(\omega) +
 f_L(\omega) }{\Gamma^2 + (\omega - \Delta)^2}
 \\
& = &
 \frac{1}{2} + \frac{1}{2 \pi} \Im \, \sum_{p=\pm}
 \Psi\left(\frac{1}{2} + \frac{\Gamma}{2 \pi T} +  i \frac{p V/2 - \Delta}{2 \pi
 T} \right) \nonumber \, ,
\end{eqnarray}
where $\Psi(x)$ denotes the psi (digamma) function. In the
simpler case of zero temperature, it simplifies to
\begin{eqnarray}
 n_\stat = \frac{1}{2} + \frac{1}{2 \pi} \sum_{p = \pm} \,
 \arctan \left(p V/2 - \Delta \right) \, .
\end{eqnarray}
Already in Eq.~(\ref{1stresult}) one easily identifies the
tunneling rate $\Gamma$ as the energy scale which governs the
approach to the steady state. Indeed, at almost all values of
other parameters the steady state (in which we can still have a
finite transport current) is established after a time of the order
$\Gamma^{-1}$. The current which is flowing during this time onto
and from the dot (depending on the initial condition) is to a
large extent the displacement current
\cite{PhysRevLett.70.4114,PhysRevB.66.195319} which is given by
the time derivative of $n(t)$,
\begin{eqnarray}                  \label{displacement}
&& I_{\text{disp}}(t) = - \frac{d n(t)}{d t}
\\ \nonumber
&=& \frac{ \Gamma \, e^{ - \Gamma t}}{\pi} \int d \omega \, \left[
f_L(\omega) + f_R(\omega)
\right] \, \\
\nonumber &\times& \frac{ \Gamma \, e^{ - \Gamma t} - \Gamma \,
\cos \left[ (\omega - \Delta) t \right] - (\omega - \Delta) \,
\sin \left[(\omega - \Delta) t \right] }{\Gamma^2 + (\omega -
\Delta)^2} \, .
\end{eqnarray}

\begin{figure}[t]
     \centering
              \includegraphics[angle=0,width=8.5cm]{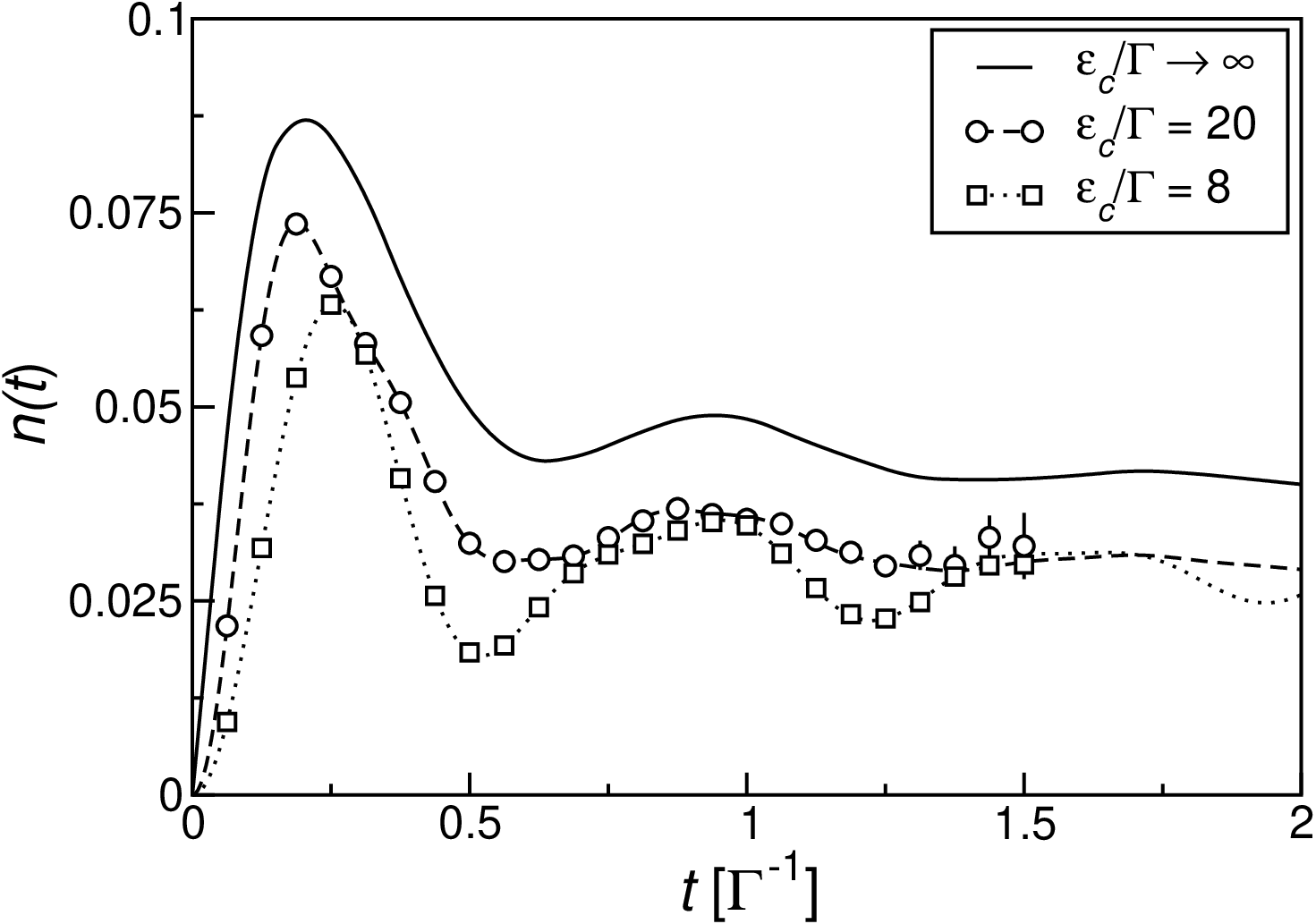}
     \caption{Population of the noninteracting dot as a function of time measured in
 units of $\Gamma^{-1}$ for $V=0$, $\Delta/\Gamma=8$ and different bandwidths: for
 $\epsilon_c/\Gamma=8$, $20$, $\infty$. Curves are the analytic results,
 symbols represent the MC simulation data (see Section \ref{MC}).}
     \label{Fig1}
\end{figure}

A surprising effect is found if the dot energy is higher than the
Fermi edges in the electrodes: on the intermediate time scale
$\Delta^{-1}$, the dot population shoots over its asymptotic
steady-state value and reaches a local maximum despite the absence
of any kind of interactions. The subsequent relaxation to
$n_\stat$ may then be either smooth, or accompanied by a number of
oscillations, as shown in Fig.~\ref{Fig1}. These are remnants of
the oscillatory behavior of the integrand in Eq.~(\ref{1stresult})
and have a period $\sim \Delta^{-1}$. In the opposite limit of a
dot lying far below the chemical potentials in the leads, one does
not observe related effects. However, as the system is
particle-hole symmetric, the analogous population oscillations are
recovered for the initially populated dot.

Next we investigate the time-dependent current. The first contribution
is essentially given by the time-dependent $n(t)$,
\begin{eqnarray}
I_L'(t) = \frac{\Gamma}{2} \, \Theta(t) \, \left[ 1- 2 n(t)
\right] \, ,
\end{eqnarray}
leading to
\begin{eqnarray}            \label{current1}
&& I_L(t) = I_{L,\stat} - \Gamma e^{- \Gamma t} \int \frac{d
 \omega}{2 \pi} \, \frac{1}{(\omega - \Delta)^2 + \Gamma^2}
 \\ \nonumber
 &\times&
 \left\{ \Gamma e^{- \Gamma t} \left[f_R(\omega) + f_L(\omega) \right]
 - \Gamma \cos[
 (\omega - \Delta) t] \left[2 f_R(\omega) + 1 \right] \right.
 \nonumber \\
 &-& \left. (\omega - \Delta) \sin [
 (\omega - \Delta) t] \left[ 2 f_L(\omega) - 1 \right] \right\}
 \nonumber \, ,
\end{eqnarray}
where the asymptotic steady-state value of the current is
\begin{eqnarray}
 I_{L,\stat} & = & \Gamma^2 \int \, \frac{d \omega}{2 \pi} \,
 \frac{f_L(\omega) - f_R(\omega)}{(\omega -
\Delta)^2 + \Gamma^2} \\
& = & \Gamma G_0 \, \mbox{Im}
\sum_{p=\pm} \, p \,
 \Psi\left(\frac{1}{2} + \frac{\Gamma}{2 \pi T} +  i \frac{p V/2 -
\Delta}{2 \pi
 T} \right) \nonumber \, ,
\end{eqnarray}
with $G_0 = e^2/h$ being the conductance quantum. The current $I_R(t)$
through the right contact is by symmetry found from $I_R(V,t) = -
I_L(-V,t)$. The difference in currents through the individual
contacts describes the charge redistribution in the system, i.e.
it must be equal to the displacement current (\ref{displacement}),
\begin{eqnarray}
 I_{\text{disp}}(t) = I_L(V,t) + I_L(-V,t) \, ,
\end{eqnarray}
which is shown by inspection. The total current through the AIM
then becomes\cite{plihal}
\begin{eqnarray}
 I(V,t) &=& \left[ I_L(V,t) + I_R(V,t) \right]/2
 \nonumber \\
 &=& \left[ I_L(V,t) - I_L(-V,t) \right]/2 \, .
\end{eqnarray}
The last two identities imply a very convenient representation of
the currents through the contacts
\begin{eqnarray}                   \label{currentdecomposition}
 I_{L,R}(t) = I(t) \pm I_{\text{disp}}(t)/2 \, .
\end{eqnarray}
Note the instantaneous current value at $t=0$,
\begin{eqnarray}                   \label{instcurrent}
 I_L(0) = \Gamma^2 \int \frac{d \omega}{2 \pi} \,
 \frac{1}{(\omega - \Delta)^2 + \Gamma^2} = \frac{\Gamma}{2} \,
 ,
\end{eqnarray}
which is independent of both voltage \emph{and} $\Delta$. In fact,
$I_L(0)$ corresponds to the current through the resonant level
system in the case of \emph{infinitely} high applied voltage.
This unphysical, \emph{instantaneous} current onset
of the left current comes as no surprise since initially every
electron in the band has the same probability of populating the
empty dot, while electrons with arbitrarily high energies allow
correspondingly fast processes. Obviously, such a situation can
never occur in any real system due to a strictly finite width of
the electrodes' conductance bands. To comply with this restraint,
we proceed with analyzing a more realistic model with finite
bandwidths.

\subsection{AIM with soft cutoff}

Depending on the material, the method of coupling to the voltage
sources, and measurement apparatus the electrode band structure
can be more complicated than for the WFB. The simplest model is a
rigid flat band with a constant DoS $\rho_0$ between $\epsilon_c'$
and $\epsilon_c$, which represent the band bottom and upper
boundary, respectively, and zero otherwise. A more sophisticated
scheme, which is more physical involves soft cutoffs at
$\epsilon_c$ and $\epsilon_c'$. Mathematically, they can be
realized as
\begin{eqnarray}               \label{cutoffscheme}
  \rho(\omega) = \frac{\rho_0}{ e^{(\omega - \epsilon_c)/\eta} + 1}
\left( 1  - \frac{1}{e^{(\omega - \epsilon_c')/\eta} + 1} \right) \, ,
\end{eqnarray}
where $\eta$ is a softening parameter. From the physical
point of view one obvious `good' value for it would be $\eta = T$,
which is the value we are using for numerical plots.\footnote{At
finite temperature the single particle levels can be considered as
broadened, having the typical width of $T$. That is why it is
natural to assume the band boundaries to be widened with $\eta
\approx T$.} The choice of two different values, $\epsilon_c$ and
$\epsilon_c'$, is deliberate. While in the equilibrium, when no
bias voltage is applied to the system, $\epsilon'_c = -
\epsilon_c$ covers the relevant physics whatever the band filling
(as long as the chemical potentials are not too close to the band
boundaries), the situation is more delicate when out of
equilibrium. Usually, the finite voltage is realized by different
chemical potentials $\mu_{L,R}$ of the two electrodes. Changing
them around the equilibrium value (we always assume the band to be
half-filled, so that in equilibrium $\mu_{L,R}=0$ in our choice of
zero point of energy) without shifting the band boundaries would
imply charging of the electrodes. In order to avoid this, one has
to shift the complete band along with the changed chemical
potential to ensure the electroneutrality, $\epsilon_c \rightarrow
\epsilon_c + \mu_{L,R}$, and $\epsilon_c' \rightarrow -\epsilon_c
+ \mu_{L,R}$. We shall see later that for voltages small compared
to the band width $2 \epsilon_c$ the changes in observables vanish
on a timescale of the order of $\epsilon_c^{-1}$. Nevertheless, in
order to be consistent we shall keep two different values
$\epsilon_c$ and $\epsilon_c'$.

A stronger DoS energy dependence is expected for a system strongly
coupled to its environment. In some cases even a DoS which
vanishes at the Fermi level may emerge. Two notable situations are
a system in the Coulomb blockade regime or a strongly interacting
low-dimensional conductor in the Luttinger liquid phase. These
are, however, systems with effectively interacting electrodes
whose treatment we postpone to a follow-up publication. From now
on, we would like to concentrate on the DoS (\ref{cutoffscheme}).

In order to solve the Dyson equation (\ref{retardedDyson}) it is
more convenient to have the retarded band GF in the time
representation. According to the prescription (\ref{theRGF}) we find
\begin{eqnarray}
 g^R(t) = \pi \rho_0 \eta \Theta(t) \, \frac{ e^{ - i \epsilon_c' t} - e^{
- i \epsilon_c t}}{ [e^{(\epsilon_c' - \epsilon_c)/\eta}-1] \sinh
\left( \pi \eta t \right)} \, ,
\end{eqnarray}
which is regular towards the limit $t=0$. From this relation one can
easily read off the retarded self-energy using Eq.~(\ref{selfenergy}).
The corresponding integral equation kernel (\ref{thekernel}) turns
out to depend only on the time difference $(t-t_2)$,
\begin{eqnarray}                            \label{fincutoffkernel}
 K(t-t_2) &=& - i \frac{ \Gamma T \, e^{ - i \Delta (t - t_2)}}{ 1 - e^{
(\epsilon_c - \epsilon_c')/\eta}}
\\ \nonumber &\times&
\int_0^{t-t_2} d \tau \, e^{ i \Delta
\tau} \, \frac{e^{ - i \epsilon_c' \tau} - e^{ - i \epsilon_c \tau}}{
\sinh (\pi \eta \tau)} \, .
\end{eqnarray}
In fact, the last integral can be expressed in terms of
hypergeometric functions. However, from a numerical point of view,
it is more convenient to work with the integral
(\ref{fincutoffkernel}) directly. In fact, writing down
the equations for the retarded GFs in the steady state case and in
the case of the sudden switching of tunneling one immediately
realizes that they are identical in the relevant time domain. In
order to calculate the time-dependent population function, one
still can use the formula (\ref{longerformula}). The necessary
off-diagonal self-energy is given by
\begin{eqnarray}          \label{fincutoffselfenergy}
 \Sigma^{<}(t) & = & - \frac{\Gamma T \, e^{- \epsilon_c/T}}{
 2 \sinh (\pi T t)} \sum_{i=R,L} \\
 & & \times \left[ \frac{e^{- i \mu_i t -
 \mu_i/T}}{(e^{- \mu_i /T } - e^{ - \epsilon_c'/T})(e^{- \mu_i/T}
 - e^{ - \epsilon_c/T})}
\right. \nonumber\\ 
& & \left.
 - \frac{e^{ - i \epsilon_c' t - \epsilon_c'/T}}{(e^{ -
 \epsilon_c'/T} - e^{-\epsilon_c/T})(e^{- \mu_i /T} - e^{ -
 \epsilon_c'/T})}
\right. \nonumber\\ 
& &\left.
 + \frac{e^{ - i \epsilon_c t - \epsilon_c/T}}{(e^{ -
 \epsilon_c'/T} - e^{-\epsilon_c/T})(e^{- \mu_i /T} - e^{ -
 \epsilon_c/T})} \right] \,\nonumber .
\end{eqnarray}
With the prerequisites (\ref{fincutoffkernel}) and
(\ref{fincutoffselfenergy}) the calculation of the time evolution
$n(t)$ as well as of the currents is a standard numerical task.
\begin{figure}[t]
\centering
\includegraphics[width=8.5cm]{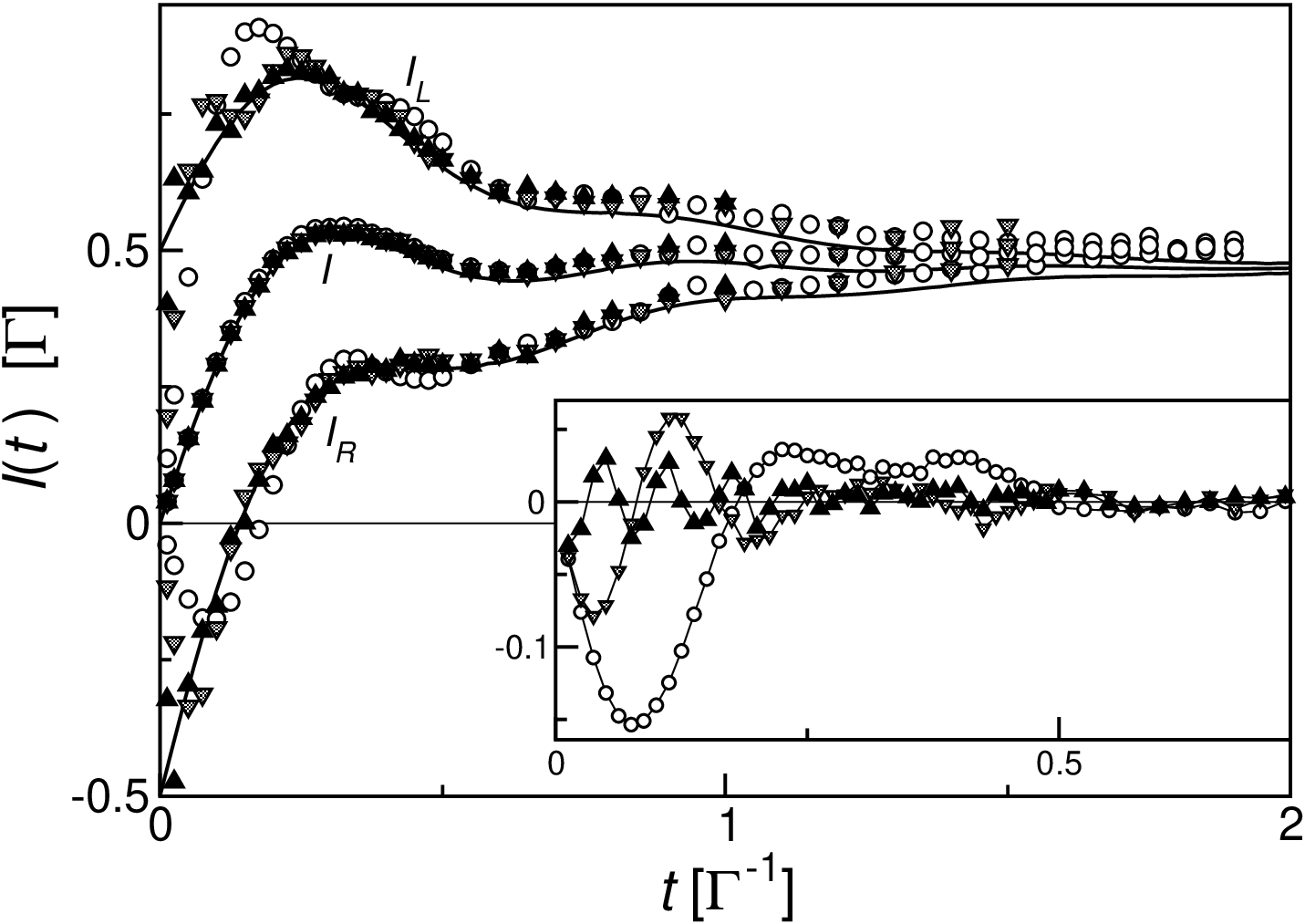}
\caption{DQMC data for the left, right, and total current $I_{L,R}(t)$, and
  $I(t)$, respectively, for $U = \Delta = 0$, $V = 20\Gamma$, and $T =
  0.4\Gamma$. Circles, triangles down, and triangles up refer to cutoff
  energies of $\epsilon_c = 20\Gamma$, $40\Gamma$, and $100\Gamma$,
  respectively. The inset shows the difference in $I(t)$ due to charge
  neutrality. The solid lines refer to the current calculated in the WFB
  limit via Eq.~(\ref{current1}).}
\label{refFig2}
\end{figure}
The results of the calculations are presented in Figs.~\ref{Fig1}
and \ref{refFig2}. The most drastic differences between the WFB model
and that with a finite bandwidth are found in the short time
behavior of the current. In contrast to the WFB prediction the
instantaneous value of currents through individual contacts is
strictly zero. Moreover, the slope (time derivative) of
$I_{L,R}(t)$ starts at zero rather than being finite. These
differences eventually vanish
 after a timescale of the order $\epsilon_c^{-1}$, so that, as expected,
the correspondence between the two calculation schemes improves.
However, the actual current behavior becomes more oscillatory and
prevents reliable simulations for too large $\epsilon_c/\Gamma >
50$. Contrary to the $I_{L,R}(t)$ currents, the total current
through the constriction is not only free of oscillations but also
shows a far better agreement with the WFB model. We conclude that the
finite bandwidth effects are contained almost completely in the
displacement component of the current (\ref{displacement}). Its
behavior is plotted in Fig.~\ref{refFig3}.

Furthermore, we find a rather small difference between the results for
systems
which preserve and neglect the electroneutrality (see inset of
Fig.~\ref{refFig2}), which only exists for
times $\sim \epsilon_c^{-1}$ and vanishes almost completely in the
steady state. We would like to point out that the maximal
deviation depicted in Fig.~\ref{refFig2} is achieved for voltages
half as large as the bandwidth. It is highly unlikely that such a
situation can ever be realized in experiments, where the maximal
voltages very seldom exceed 5\% of $2\epsilon_c$. Therefore from
now on we refrain from implementing the electroneutrality
requirement in our analysis.

\begin{figure}[t]
\centering \vspace*{0.25cm}
\includegraphics[width=8.5cm]{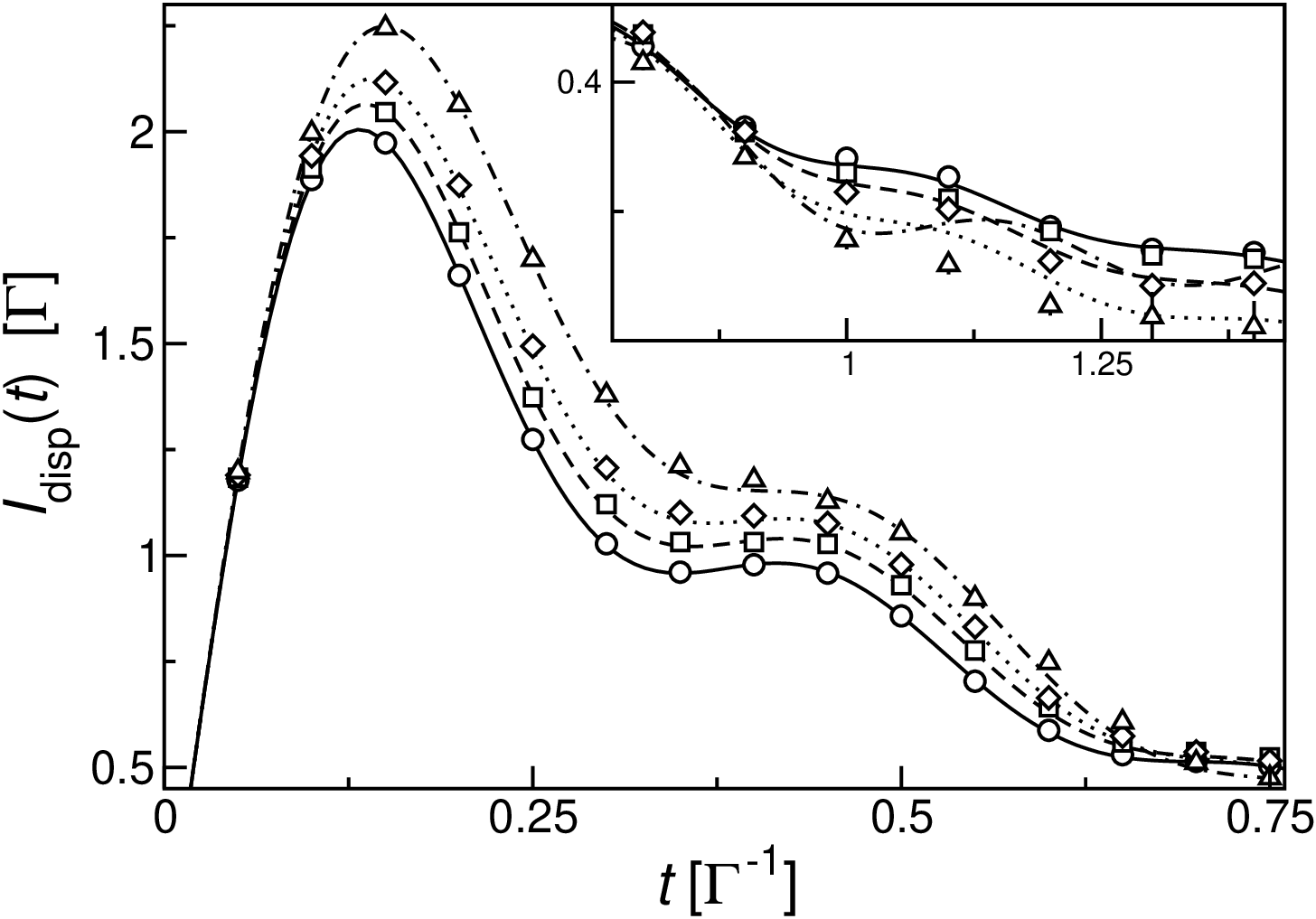}
\caption{DQMC data for the displacement current $I_\text{disp}(t)$ (open
  symbols) and results according to the perturbation theory (lines)
  for $V=0$, $\epsilon_c = 20\Gamma$, and $T=0.4\Gamma$.
Circles, squares, diamonds, and triangles (solid, dashed,
dotted, and dotted-dashed lines) refer to $U = -2\Delta = 0$, $\Gamma$,
$2\Gamma$, and $4\Gamma$, respectively.}
\label{refFig3}
\end{figure}

\section{Perturbation theory in interaction}   \label{perttheory}

As a next step, we investigate the change of the dot transient
dynamics due to the finite Coulomb repulsion, which is described
by the term (\ref{H_U}) in the Hamiltonian. In the regime where
$U$ is small compared to the other energy scales, we can employ a
perturbative expansion which we truncate after the first order.
Note that as the interaction involves electrons of opposite spins,
we have to keep track of the spin indices henceforth.

In order to calculate the time-dependent dot population
$n_\sigma(t)$, we start from Eq.~(\ref{nddef}) and expand the dot
GF to first order in $U$. Discarding all disconnected
diagrams, this leads to
\begin{equation}
    D^{(1)}_\sigma(t,t') =
    U \int_C ds\ n_\sigbar(s) D_\sigma(t,s) D_\sigma(s,t') \, .
\end{equation}
The superscript denotes the first order in $U$, while the GFs on
the right hand side and the particle density $n_\sigbar(s)$ are
the respective functions for the noninteracting case. The time
integration runs along the Keldysh contour $C$. The lesser GF can
be expressed in terms of the advanced and retarded GFs and reads
\begin{eqnarray}
    D^{(1)<}_\sigma(t,t') & = &
    U \int_{-\infty}^\infty ds\ n_\sigbar(s)
    \big[ D^R_\sigma(t,s) D^{<}_\sigma(s,t') \nonumber \\
    & + &
    D^{<}_\sigma(t,s) D^A_\sigma(s,t') \big] \, .
\end{eqnarray}
Both components can be combined by using the complex conjugation
properties
\begin{eqnarray}
    D^A_\sigma(t,t') & = & [D^R_\sigma(t',t)]^* \nonumber \, ,\\
    D^{<}_\sigma(t,t') & = & - [D^{<}_\sigma(t',t)]^* \, .
\end{eqnarray}
Initially, before the coupling to the leads is switched on, the
dot is assumed to be empty, which means that $n_\sigbar(s) = 0$
for $s < 0$. Therefore, we can rewrite the first order correction
to the dot occupation number as
\begin{equation}
    n^{(1)}_\sigma(t) = 2\ U \Im[r_\sigma(t)]\ ,
\end{equation}
where
\begin{equation}\label{f_sigma}
    r_\sigma(t) = \int_0^t ds\ n_\sigbar(s) D^R_\sigma(t,s)
D^{<}_\sigma(s,t)
\end{equation}
depends only on properties of the noninteracting system and is
thus accessible. In the WFB limit, this calculation can mostly be
done analytically using the functions given in Eqs.~(\ref{gK_WB})
and (\ref{DR_flat}).

In order to keep the derivation simple, we shall investigate the case
$\Delta = 0$ in equilibrium and at zero
temperature ($V=T=0$). The change in the asymptotic value
can most easily be accessed, because the usual perturbation theory
in the frequency domain can be employed. One finds the following
correction to the dot occupation,
\begin{equation}
 n_{\sigma,\stat}^{(1)} = - i U n_{\sigbar,\stat}^{(0)} \int
\frac{d\omega}{2\pi}
 D^{<}_\sigma(\omega) \left[ D_\sigma(\omega) - \widetilde{D}_\sigma(\omega)
 \right]\ .
\end{equation}
Using the well-known expressions for the dot GFs
$D^{ij}_\sigma(\omega)$, and the fact that the unperturbed
stationary dot occupation is given by
$\smash{n^{(0)}_{\sigma,\stat}} = 1/2$, one finds
\begin{equation}
 n_{\sigma,\stat}^{(1)} = -\frac{U}{2\pi\Gamma}\ ,
\end{equation}
which is a simple reduction of the stationary value due to the
Coulomb repulsion. In a next step, we shall find out how this
stationary value is approached. For this purpose, we evaluate the
function (\ref{f_sigma}) which contains the unperturbed
time-dependent occupation number. For our choice of parameters,
this function can be read off from Eq.~(\ref{1stresult}):
\begin{equation}
 n^{(0)}_\sigbar(t) = \frac{1}{2} \left( 1 - e^{-2 \Gamma t} \right) \,
.
\end{equation}
Hence, in equilibrium the dot occupation without Coulomb
interaction saturates exponentially on a time-scale $\Gamma^{-1}$. Moreover, we need the lesser dot GF which can be
derived from Eq.~(\ref{longerformula}). One finds
\begin{align}
 D^<_\sigma(s,t)
& =
 -\frac{i \Gamma}{\pi} \int_{-\infty}^0 \frac{d\omega}{\Gamma^2 + \omega^2}
 \\
\times&
 \left[ e^{-i\omega(s-t)} - e^{-\Gamma s} e^{i\omega t} - e^{-\Gamma t}
 e^{-i\omega s} + e^{-\Gamma(s+t)} \right] \, .\nonumber
\end{align}
For $\Delta = 0$, the retarded GF (\ref{DR_flat}) becomes purely
imaginary and the only imaginary part in Eq.~(\ref{f_sigma})
arises from the lesser GF $D^<_\sigma(s,t)$. Therefore, we only
need to evaluate the imaginary part of the $\omega$-integrals.
With the definition\cite{gradsteyn75}
\begin{eqnarray}
 z(t)
& := &
 \Im \int_{-\infty}^0 d\omega \frac{e^{i \omega t}}{\Gamma^2 + \omega^2}
 \\
& = &
 \frac{\sgn(t)}{\Gamma} \left[ \text{Chi}( \Gamma |t| ) \sinh( \Gamma |t| )
 - \text{Shi}( \Gamma |t| ) \cosh( \Gamma |t| ) \right],
\nonumber
\end{eqnarray}
where $\text{Shi}(x)$ and $\text{Chi}(x)$ denote the hyperbolic sine and cosine
integral functions, respectively, we obtain
\begin{eqnarray}                            \label{nt_int}
 n_\sigma^{(1)}(t)
& = &
 \frac{2 U \Gamma}{\pi} \int_0^t n_\sigbar^{(0)}(s) e^{-\Gamma(t-s)}
 \\
& \times &
 \left[ z(t-s) - e^{-\Gamma s} z(t) + e^{-\Gamma t} z(s) \right]\ .
 \nonumber
\end{eqnarray}
This integral is evaluated numerically for arbitrary $t$. As the
integrand is regular towards $s, t \rightarrow 0$, one concludes
that the time derivative vanishes for small times and the correction
due to the Coulomb interaction only sets in gradually. This is
understandable since we assumed the dot to be initially unoccupied
and the Coulomb interaction can only have an effect once a finite
population has been established on the dot.

\begin{figure}[t]
     \centering
             \includegraphics[angle=0,width=8.5cm]{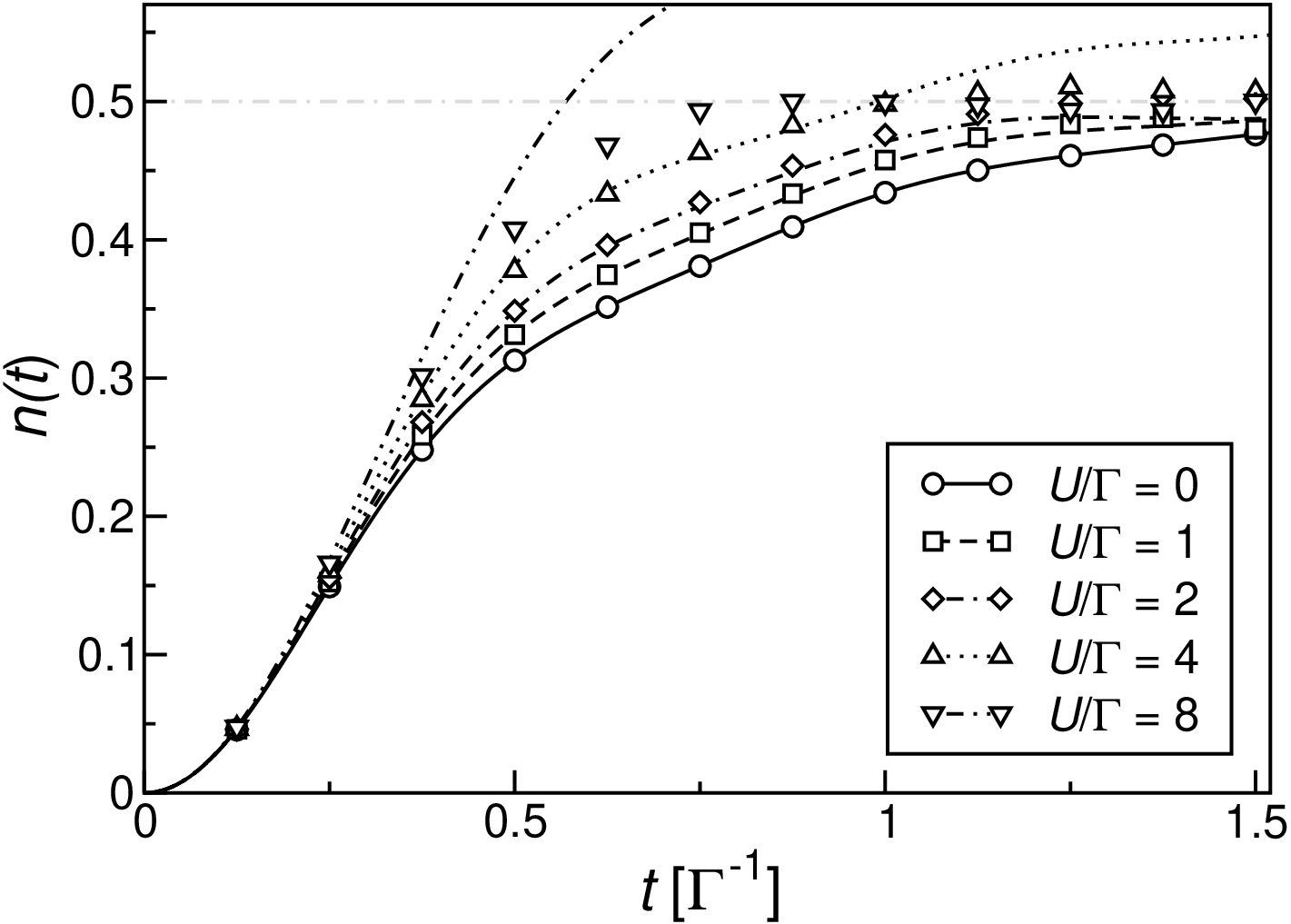}
     \caption{Time-dependent dot occupation for different interaction
     strengths $U$
     ($V=0$, $\Delta=-U/2$, $\epsilon_c/\Gamma=10$, $T=0.2\Gamma$).
     Curves show the results from the first order perturbation
calculation in $U$. The symbols represent the MC data.}
     \label{refFig4}
\end{figure}

The previous calculations for $V = \Delta = T = 0$ clearly present
an oversimplified picture. Although the qualitative statements
remain correct, additional energy scales due to finite
temperature, asymmetry, voltage, and bandwidth render the whole
picture more complicated. In these cases, however, an
all-numerical scheme has to be used. We chose to solve the
integral equation (\ref{retardedDyson}) by discretizing the time
axis and thus translating it into a matrix equation.

In order to investigate the limit of the approximation in $U$, we
compared the perturbative corrections to MC results which are
exact for any interaction strength. In Fig.~\ref{refFig4}, we plot
the time-dependent occupation probability for a single spin
orientation in the noninteracting model and for a relatively small
interaction $U$. While for small times, the graphs coincide within
numerical accuracy, deviations become visible after a time of
order $\Gamma^{-1}$. This is not surprising because in order for
the interaction term to be fully operational a finite dot
population is necessary. This process requires a time of the
order $\Gamma^{-1}$.

\begin{figure}[t]
     \centering
             \includegraphics[angle=0,width=8.5cm]{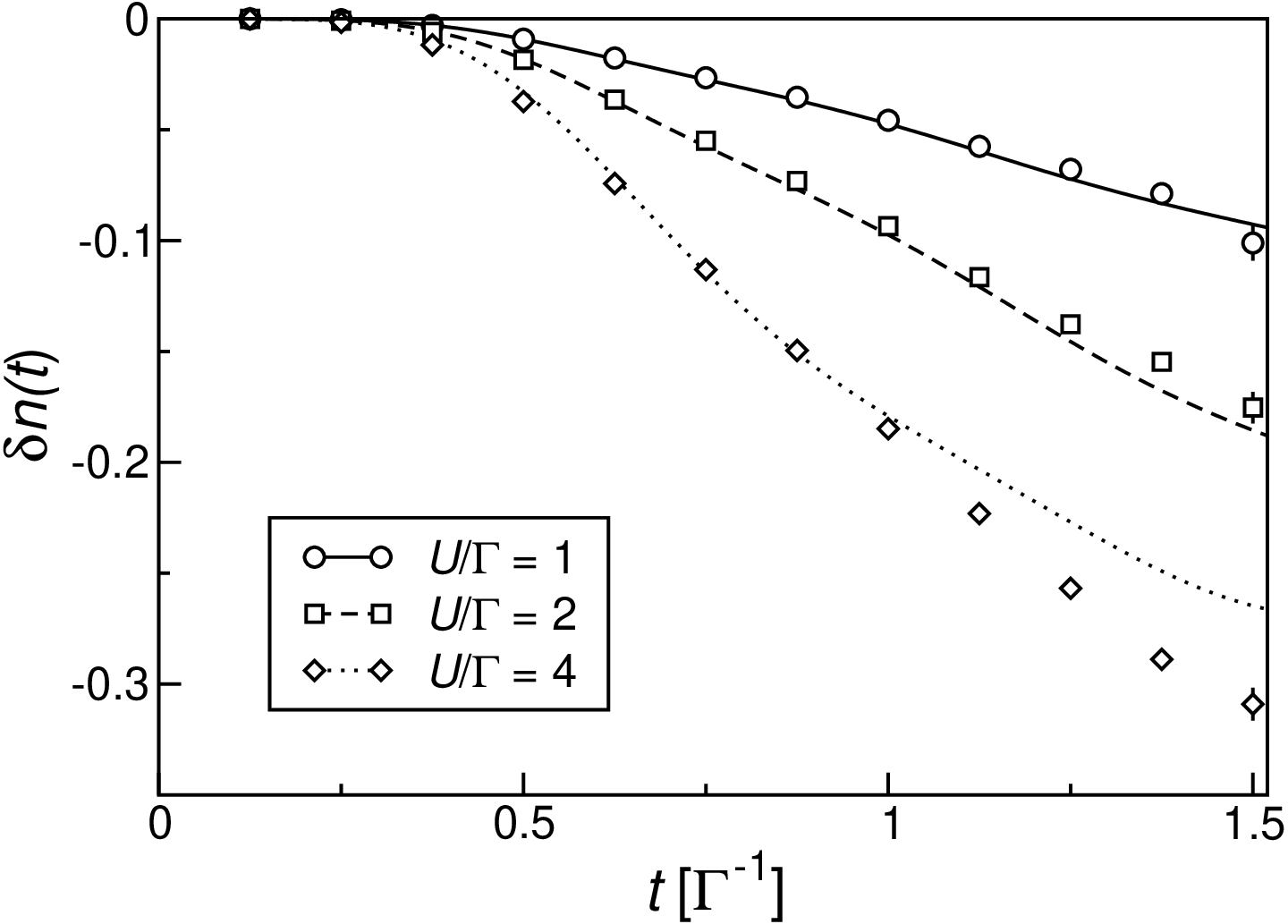}
     \caption{Change of the time-dependent dot occupation due to
interactions, $\delta n_\sigma(U) = n_\sigma(U)-n_\sigma(U=0)$, for several values of
the interaction strength. $V=0$, $\Delta=-U/2$,
$\epsilon_c/\Gamma=10$, $T=0.2\Gamma$. Curves show the results from the first order perturbation
calculation in $U$. The symbols represent the MC data.}
     \label{refFig5}
 \end{figure}

In order to investigate the interaction effect more closely, we plot in
Fig.~\ref{refFig5} the
interaction correction for several values of $U$. One notices a good
agreement up to times of order $\Gamma^{-1}$
even for interaction strengths as large as $U=\Gamma$, which is actually beyond the range of
the perturbation theory. In this regime, even
an expansion up to second order in $U$, albeit feasible in
principle, would not lead to a more reliable result. For this
reason, we only consider the first order in $U$.

While the effect of the Coulomb interaction on the displacement
current can be calculated by differentiating Eq.~(\ref{nt_int})
with respect to time, we still have to investigate the time
dependence of the average current. As we argued previously, its
dependence on the electronic bandwidth is very small, so we shall
do this analysis in the WFB limit. The calculation starts again
from Eqs.~(\ref{Ip1}) and (\ref{Ip2}), where we have to use the
first order expansions of the dot GFs. Hence the expression for
the current across, say, the left lead, is given by a sum of two
terms $\smash{I^{(1)}_L = I'^{(1)}_L(t) + I''^{(1)}_L(t)}$, where
the first one is proportional to the dot occupation
\begin{equation}                  \label{tok1}
    I'^{(1)}_{L\sigma}(t)
=
 \Gamma \, \Theta(t) \, n_\sigma^{(1)}(t)\ .
\end{equation}
The derivation of the second contribution involves the complete
set of dot GFs, $D^K$, $D^R$ and $D^A$. A straightforward
calculation then leads to the following result which is valid at
zero temperature,
\begin{eqnarray}                  \label{tok2}
 I''^{(1)}_{L\sigma}(t)
& = &
 \frac{e U \Gamma}{\pi} \int_0^t ds\ \frac{e^{-\Gamma(t-s)}}{t-s} \\
& \times & \left\{ \frac{1}{2} - \cos[| V/2 - \Delta | (t-s)] \right\}
 \int_s^t ds'\ n_\sigbar^{(0)}(s'). \nonumber
\end{eqnarray}
This integral can be calculated numerically using the known
zero-order dot population $\smash{n^{(0)}_\sigbar(t)}$. The result
is compared with the MC data in Fig.~\ref{refFig6}. In contrast to
the dot occupation, the agreement between perturbation theory and
numerically exact MC simulations for the current degrades rapidly
towards stronger interaction, see Fig.~\ref{refFig6} for
$U=8\Gamma$. The steady state value of the current is considerably
smaller than that for weak interactions. However, this does not
contradict the conventional Kondo picture, where the conductance
is enhanced, since in our case the voltage is higher than the
interaction strength. Another feature is the extremely weak
oscillations in relation to the small $U$ case, which can be
interpreted as a precursor of the Kondo physics as discussed in
[\onlinecite{goker07}].

\section{Monte Carlo simulations at arbitrary interaction
strength}                           \label{MC}

To go beyond the first order perturbation theory, we employ a
numerically exact diagrammatic Monte Carlo scheme. The method is a
generalization of the algorithm proposed in
Ref.~[\onlinecite{Muehlbacher08}], or -- equivalently -- the
Keldysh implementation of the diagrammatic impurity solver of
Ref.~[\onlinecite{Werner06}]. Here, we only outline the basic
concepts of the algorithm and refer to
Refs.~[\onlinecite{Werner06,Muehlbacher08}] for further details;
the specific aspects of dealing with the electron-electron
correlations in the real-time simulations will be presented in a
forthcoming publication.

The main idea is to evaluate the expectation value $\langle
O(t)\rangle={\rm Tr}_\text{dot,lead} [\varrho_0^{\text{dot,lead}}
e^{itH}O e^{-itH}]$ by stochastically sampling a perturbation
expansion in the tunneling term $H_T$. To this end, we employ an
interaction representation in which the time evolution along the
Kadanoff-Baym contour $0\rightarrow t \rightarrow 0$ is determined
by $H_\text{loc} +H_0=H_\text{dot}+H_U+H_0$ and rewrite the time
evolution operators $e^{\pm
  itH}$ as (anti-)time ordered exponentials
\begin{eqnarray}
\langle O(t) \rangle
&=&
{\rm Tr}_\text{dot,lead}\Big[
\varrho_0^{\text{dot,lead}} \tilde T e^{i\int_0^t ds H_T(s)} O(s)\nonumber\\
&& \hspace{16mm}
T e^{-i\int_0^t ds H_T(s)} \Big] \,,
\end{eqnarray}
with $O(s) = e^{it(H_\text{loc} +H_0)} O
e^{-it(H_\text{loc}+H_0)}$ (and $H_T(s)$ accordingly). The
exponentials are then expanded into a power series, which allows
to trace out the electrode degrees of freedom in an exact manner.
At perturbation order $N$ (for given spin), this yields a
determinant of an $N \times N$ matrix whose elements are
determined by the self-energy and the times at which the tunneling
events occur. The configuration space consists of all possible
sequences of dot creation and annihilation operators on the
Kadanoff-Baym contour, and the MC sampling proceeds through local
updates of these operator sequences (insertion/removal of pairs of
creation and annihilation operators, or shifts of the operator
positions). We use both a continuous-time implementation (CTQMC)
as well as one which utilizes a discretization scheme for the
real-time axis to speed up the sampling process (DQMC). While the
continuous-time approach is completely free of systematic errors,
we kept discretization effects below statistical errors in the
discrete time implementation as well, by using a fine mesh of
$10^3$ points along the real-time axis.  From the simulation, we
obtain the time-dependent dot population $n(t)$ as well as
$I_{L,R}(t)$, $I_{\text{disp}}(t)$, and the total current $I(t)$.

\begin{figure}[t]
\centering
\includegraphics[width=8.5cm]{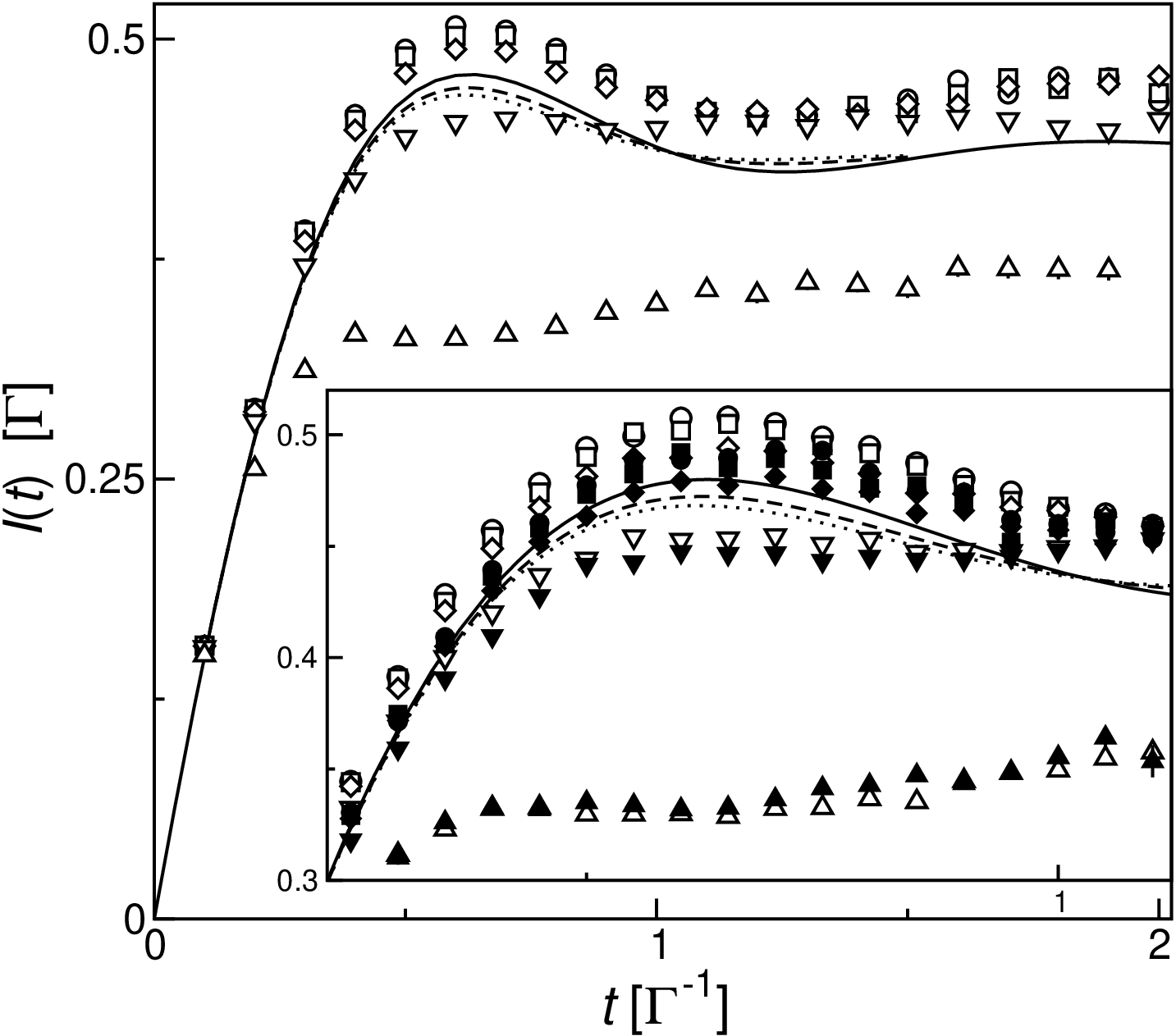}
\caption{DQMC data for the total current $I(t)$ (symbols) and
results according to Eqs.~(\ref{tok1}) and (\ref{tok2}) (lines)
for $V=10\Gamma$, $T=0$, and $\epsilon_c = 10\Gamma$ and
$40\Gamma$ (open and filled symbols, respectively). A WFB was
assumed for the first order perturbation calculation. Circles,
squares, diamonds, and triangles (facing down and up) refer to $\Delta = - U/2$ and $U=
0$, $\Gamma$, $2\Gamma$, $4\Gamma$, and $8\Gamma$, respectively, 
while solid, dashed, and dotted lines refer to U = 0, $\Gamma$, and 
$2\Gamma$.} \label{refFig6}
\end{figure}

While this simulation approach can handle arbitrary interaction strengths, it
suffers from a dynamical sign problem \footnote{In imaginary time, the
  diagrammatic Monte Carlo simulation of the AIM does not encounter a sign
  problem. All the (fermionic) sign cancelations between crossing and
  non-crossing diagrams can be absorbed into the determinant of self-energies
  (hybridization functions), and the imaginary time evolution
  $e^{-H_\text{loc}\tau}$ gives a real contribution to the weight. In real-time
  diagrammatic Monte Carlo, both the self-energies and the time evolution
  operator are complex.}  which becomes severe at long times or for large
bandwidth. We find that the error bars grow exponentially with
average perturbation order, and thus exponentially with time $t$.
In the noninteracting model, which factorizes into spin-up and
-down components, only one spin species needs to be simulated.
This reduces the average perturbation order by a factor of two and
allows us to simulate a time interval which is about twice as long
as in an interacting model (in contrast to imaginary time, the
perturbation order is essentially independent of interaction
strength). Our simulations of the interacting AIM can reach the
stationary state for certain parameters (cf.~Fig.~\ref{refFig6}), but
not yet for in the general case. We note in passing that in
principle, the inclusion of a phonon background coupling to the
interacting dot is straightforward within the path-integral
framework of Ref.~[\onlinecite{Muehlbacher08}], or using the
method proposed in Ref.~[\onlinecite{Werner07}].

\section{Discussion and conclusions}
\label{discussion}
After switching on the tunneling, the dynamics of the AIM exhibits a
surprisingly rich transient behavior. The reason can be found in the abundance
of different energy scales. While in the steady state, all parameters like
temperature $T$, voltage $V$, contact transparency $\Gamma$, and dot parameters
-- the bare energy $\Delta$ and the interaction strength $U$ -- are known to be
decisive for the stationary values of the transport current as well as the
population probability of the dot, in the opposite limit of intermediate and
especially short times the dynamics is dominated by the influence of $\Gamma$
and $\Delta$. In fact, in both the interacting and the noninteracting case the
typical timescale at which the steady state is reached is of the order
$\Gamma^{-1}$.  In the noninteracting case it is clearly visible that the
approach towards steady state is exponential, see Eqs.~(\ref{1stresult}) and
(\ref{current1}).  On the other hand, a nonzero detuning $\Delta$ or a finite band cutoff lead to superimposed oscillations of the observables' time evolution.

The most interesting behavior is encountered at very short timescales. It turns
out that the rather simple WFB theory fails to give meaningful results here,
predicting e.g. an instantaneous finite value for the currents through the
individual contacts (\ref{instcurrent}). This unphysical picture can only be
corrected by considering a more realistic model for the electrodes featuring a
finite bandwidth $\epsilon_c$, which then dominates the short time dynamics of
the system. It slows down the onset of the current and dot population and thus
quenches their time derivative to much smaller values than for an infinite
cutoff. Only after a timescale of the order of $\epsilon_c^{-1}$ do current and
dot population approach the values of the WFB model.

However, this comes as no surprise since for a system with a finite range of
allowed excitations $W$ (in our case the electrodes with $W \sim
\epsilon_c$), the uncertainty principle demands that the reaction to any
instantaneous perturbation (switching on of tunneling) has to take place on a
finite timescale $\sim W^{-1}$. Furthermore, the MC simulations reveal fast
oscillations in the currents through the individual contacts, whose wavelength and
amplitude decreases with increasing $\epsilon_c$.

In contrast to the currents through the electrodes and the diplacement current,
however, the total current through the system is only weakly affected by the bandwidth.
According to the numerical results, even for bandwidths only
twice as large as the voltage, the current follows the analytical results
for the WFB with high accuracy. This stems from the fact that the displacement
current (\ref{displacement}) monitors the
redistribution of charge across the electrodes, but it is not responsible
for any net charge flow; on the other hand, the cutoff effects are almost
completely due to the charge redistribution. Therefore, it appears natural to
expect the total current to be only weakly affected by the explicit value of
$\epsilon_c$.
We find virtually no
influence on the long-time dynamics for realistic voltage/bandwidth quotients.
The maximal deviations with respect to the WBF limit is again achieved during
times $\sim \epsilon_c^{-1}$.

For an initially empty dot the effects of a finite Coulomb
interaction become visible only after a timescale $\Gamma^{-1}$. This can
be rationalized by observing that the same timescale is necessary to build up a
dot population large enough to be affected by electronic correlations.
Furthermore, we find that the quality of the approximation by lowest order
perturbation expansion is remarkably good (even for intermediate $U$) up to
$t\Gamma\approx 1$, which is about a factor of three smaller than the time required to reach the steady state.

This shows that similarly to the Kondo case of
Ref.~[\onlinecite{PhysRevLett.83.808}], the full interaction effects take a time of several $\Gamma^{-1}$
to develop.
Another interaction effect is the suppression of both the dot population and the current. While the first effect is quite natural, the second one
is seemingly at odds with the common wisdom that at sufficiently low
temperatures, due to the Kondo effect, the transport properties of the system
must approach the unitary limit of a perfectly resonant level, see e.g.
Ref.~[\onlinecite{kaminski}]. Our results do not allow to see this kind of physics for
two reasons: (i) the steady state is not yet fully established, (ii) the
applied voltage is rather large and therefore has the potential to destroy the Kondo effect even in the
steady state regime.

To conclude, we presented a theoretical treatment of transient effects in an AIM biased
with a finite voltage after a sudden switching on
of the tunneling. Using exact analytical solutions and
perturbation theory as well as dedicated numerical schemes (MC) we identified
different regimes in the time evolution of the currents and the dot's
population probability and related them to the parameters of the system. Special
attention has been paid to the influence of electron-electron interactions on the dot
and the bandwidth of the electrodes.


\acknowledgments T.L.S.~is financially supported by the Swiss NSF and the
NCCR Nanoscience. A.K.~is supported by the DFG grant No.~KO-2235/2
(Germany). L.M.~acknowledges computational resources from the Black Forest Grid at the university of Freiburg. P.W.~was supported by NSF-DMR-0705847. The CTQMC calculations were run on the Hreidar cluster at ETH Zurich.

\bibliography{bibfile}

\end{document}